\documentstyle[twoside,fleqn,epsfig,rotating,espcrc2]{article}

\title{
Implications of HERA results for very high energy cosmic 
ray physics\thanks{Invited talk given at the Xth Int. Symposium on Very
High Energy Cosmic Ray Interactions, Gran Sasso, July 12-17, 1998.}
\vspace*{-3.5cm}
\begin{flushright}
\normalsize 
preprint BA--98--45\\
October 1998
\end{flushright}
\vspace*{2.5cm}
}

\author{Ralph~Engel
\address{University of Delaware, Bartol Research Institute, 
Newark, DE 19716 USA, e-mail: eng@lepton.bartol.udel.edu}}

\begin{document}

\begin{abstract}
Experimental results obtained with the HERA collider and recent 
progress in their theoretical interpretation are reviewed.
After a short introduction to HERA physics, deep inelastic scattering
and photoproduction are discussed as (virtual) photon-proton scattering.
It is shown that the measurement and theoretical understanding
of both photoproduction as well as low-$x$ deep inelastic scattering 
are essential for obtaining reliable high energy extrapolations 
within hadron-hadron interaction models. Limitations of the 
predictive power of hadron interaction models due to the interplay of
perturbative QCD and unitarity effects are discussed.
\end{abstract}

\maketitle

\section{Introduction}

It is very difficult to compare data measured at the HERA collider
directly to data
obtained in cosmic ray experiments. Almost all cosmic ray experiments deal
with final state particles (hadrons, electrons, photons, muons, or
neutrinos) which are produced in interactions of primary cosmic ray
(CR) particles
with air. Furthermore, the experiments are very sensitive to the projectile
fragmentation region. In the contrary, modern collider experiments focus
on particle production in the central pseudorapidity range in the
projectile-target center-of-mass system (CMS). They either
deal with leptons, (anti)protons or heavy nuclei as primaries. 
Nevertheless, it is
possible to relate particle production processes in CR
interactions to those studied in collider experiments. 

Since the number of final state particles is often very large, efficient
approximations as well as simulation techniques have to be applied to
derive conclusions from measured data or to study theoretical predictions.
One of the most powerful technique is the realization of theoretical
models in Monte Carlo event generators.
Such event generators allow us to study hadron production at
colliders as well as in CR interactions.

Unfortunately, our current understanding of hadronic interaction
processes is rather limited. Therefore, these event generators can only
be built by combining theoretical predictions with phenomenological
models and parametrizations and have to be tuned by comparing
their predictions to fixed target and collider measurements. 
Considering the underlying theory and models entering such 
Monte Carlo programs, 
almost all data measured at HERA are relevant 
to our understanding of very high energy cosmic ray interactions. 
Examples include data on\\
(i) limits on physics beyond the Standard Model,\\
(ii) parton densities,\\
(iii) reliability and range of applicability of perturbative QCD and
various approximation schema,\\
(iv) transition between soft and hard physics (perturbative and 
non-perturbative regimes),\\
(v) forward hadron production, and \\
(vi) heavy flavour production.

It is beyond the scope of this work to discuss all these data.
Instead, we shall focus on the implications of the low-$x$ structure
function measurements \cite{Aid96b,Derrick96c,Adloff97a,Breitweg97b}
and recent results on forward jet \cite{Breitweg98a,Adloff98a} and 
leading baryon production \cite{Garfagnini98a,Schmidke98a}.
Where possible, comparisons to predictions of 
Monte Carlo models currently used in
the analysis of CR data are included.

After a short introduction to the HERA collider, deep inelastic
scattering (DIS) and parton densities at low $x$ are discussed in
Sec.~\ref{part3}. In Sec.~\ref{part4} DIS is interpreted as (virtual)
photon-proton scattering and data on leading jet and baryon production are
presented.
Finally, in Sec.~\ref{part5} a summary is given.


\section{The HERA collider\label{part2}}

The electron-proton storage ring HERA at DESY, Hamburg (Germany) went
into operation in 1992. At HERA, 27.6 GeV positrons or electrons are 
collided on 820 GeV protons.
The lepton\footnote{Thereafter the term 'lepton' is used to refer either
to electrons or positrons accelerated at HERA.}-proton 
c.m.\ energy is $\sqrt{s}_{ep} \approx 300 $ GeV which
corresponds to a lepton energy of $E_{\rm lab} \approx 47 \times
10^{12}$ eV in the proton rest frame. 

\begin{figure}[!htb]
\centerline{\epsfig{figure=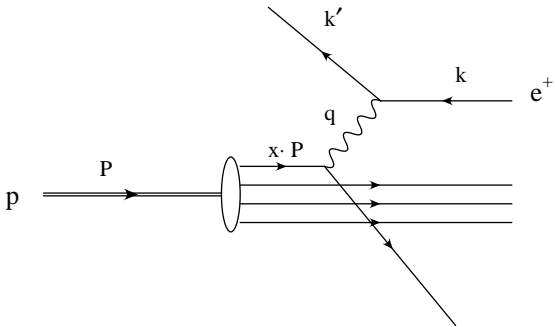,width=7.5cm}}
\vspace*{-5mm}
\caption{Kinematics of positron-proton scattering in Breit frame.
\label{fig-1}
}
\end{figure}

The typical scattering process
studied at HERA is sketched in Fig.~\ref{fig-1}. 
The beam lepton
scatters off the proton by exchanging a photon (with a small cross
section, the exchanged particle might be also a $W^\pm$, $Z$ gauge
boson). The kinematics of the reaction can be characterized by the following 
quantities
\begin{itemize}
\item photon virtuality: $Q^2 = -q^2 = -(k-k^\prime)^2$
\item Bjorken's scaling variable: $x = Q^2/(2 P\cdot q)$
\item inelasticity: $y = P\cdot q /(P\cdot k)$.
\end{itemize}
There are only two independent variables, we shall use $x$ and $Q^2$.
In the proton infinite-momentum frame, $x$ denotes the momentum fraction
of the proton carried by the struck quark.

\begin{figure}[!htb]
\centerline{\epsfig{figure=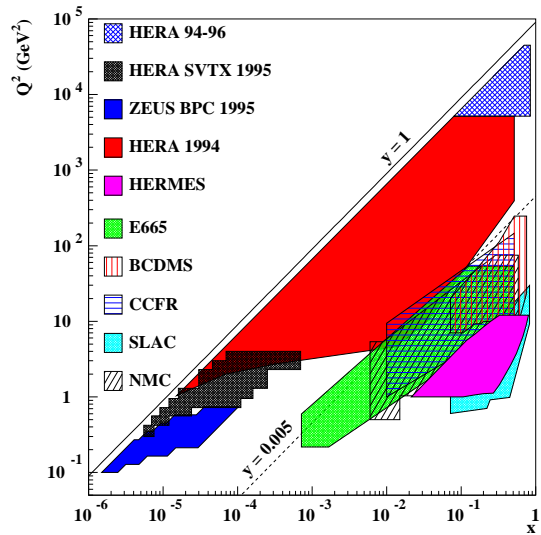,width=7cm}}
\vspace*{-5mm}
\caption{Kinematic region of DIS covered by experiments up to now
\protect\cite{Schneekloth98a}.
\label{fig-2}
}
\end{figure}

In Fig.~\ref{fig-2}
the kinematic range for DIS accessible by various fixed-target 
experiments and
HERA is shown. 
The HERA measurements extend the previously explored kinematical range
by about two orders in magnitude in $x$ and $Q^2$.


\section{Structure function and low-$x$ parton densities\label{part3}}

The structure functions $F_2$ and $F_L$ are defined by the differential
lepton-proton cross section
\begin{eqnarray}
\frac{d^2\sigma_{ep}}{dx dQ^2} &=& \frac{2\pi\alpha^2}{Q^4 x}
\bigg[ (1+(1-y)^2) F_2(x,Q^2)
\nonumber\\
& &-y^2 F_L(x,Q^2) \bigg]\ ,
\end{eqnarray}
where $\alpha$ denotes the fine structure constant.
For most applications the longitudinal structure function $F_L$ can be 
neglected since $y$ is sufficiently small.

\begin{figure}[!htb]
\centerline{\epsfig{figure=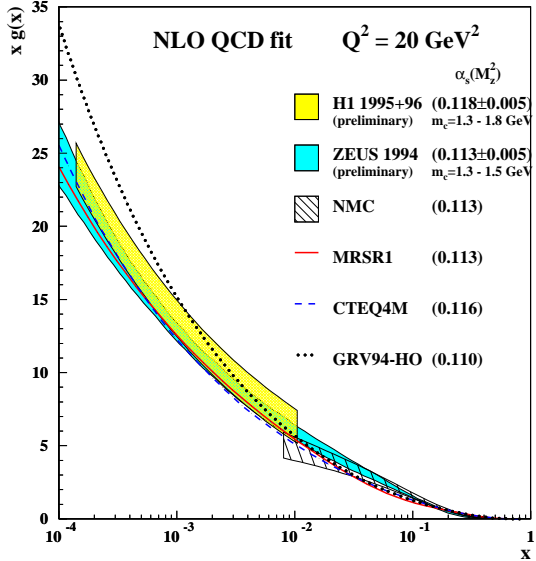,width=7cm}}
\vspace*{-5mm}
\caption{Gluon density at low $x$ and $Q^2=20$ GeV$^2$. Shown are H1,
ZEUS and NMC data together with several parton density parametrizations.
\label{fig-3}
}
\end{figure}

In leading order perturbative QCD the structure function
$F_2$ is related to the quark densities $q, \bar q$ in the proton by
\begin{equation}
 F_2(x,Q^2) = x \sum_{i} e_i^2 \left( q_i(x,Q^2) +
\bar{q}_i(x,Q^2)\right),
\end{equation}
where  $e_i$  is fractional charge of the quark $i$.
The Dokshitzer-Gribov-Lipatov-Altarelli-Parisi (DGLAP) equations
\cite{Gribov72a,Dokshitzer77a,Altarelli77a}
predict the $Q^2$ dependence of the quark and gluon densities (for large
$Q^2$ and not too small $x$)
\begin{eqnarray}
\frac{\partial}{\partial \ln Q^2}
\left(\begin{array}{c}
q(x,Q^2)\\
g(x,Q^2)
\end{array}
\right) & &
\nonumber\\
& &\hspace*{-3.5cm} = \frac{\alpha_s(Q^2)}{2 \pi}
\left(\begin{array}{cc}
P_{qq} & P_{q g} \\
P_{ g q} & P_{ gg}
\end{array} \right)
\otimes \left(\begin{array}{c}
q(x,Q^2)\\
g(x,Q^2)
\end{array}
\right) .
\label{DGLAP}
\end{eqnarray}
Here $P_{ij}$ represents the splitting kernel describing the 
probability of finding a daughter parton $i$ in the parent parton $j$ 
(e.g.\ $\sum_k P(j \rightarrow i, k$)) and we have used
$\alpha_s = g_s^2/(4\pi)$
with $g_s$ being the strong coupling constant.
In the limit $\ln(1/x)$,~$\ln(Q^2/\Lambda^2) \rightarrow \infty$ 
(double leading-logarithmic approximation) the DGLAP
equations predict a steeply rising gluon density $g(x,Q^2)$
\begin{eqnarray}
x g(x,Q^2) &\sim& \exp\left[ \frac{48}{11-\frac{2}{3}n_f}
\ln \frac{\ln \frac{Q^2}{\Lambda^2}}{\ln \frac{Q_0^2}{\Lambda^2}} \ln
\frac{1}{x} \right]^{\frac{1}{2}}
\label{DGLAP-gluon}
\\
&\sim& \frac{1}{x^{0.4}}, 
\label{double-log}
\end{eqnarray} 
where $\Lambda$ denotes the QCD renormatization scale and $n_f$ 
is the number of quark flavours.
Since the gluon and sea quark densities are closely related
(see Eq.~(\ref{DGLAP})), 
a rise of $F_2$ is expected for decreasing $x$. 
%
%
This has been confirmed by H1 and ZEUS measurements. Experimentally, 
the gluon density
can be estimated, for example, from the scaling violation of $F_2$ ($Q^2$
dependence) or from the diffractive $J/\Psi$ production cross section.
In Fig.~\ref{fig-3} the gluon density is shown for $Q^2 = 20$ GeV$^2$
together with several theoretical predictions.
Currently, data on quark and gluon densities are
available for $x$ as low as $10^{-4}$ and $Q^2 \approx 20$ GeV$^2$. 
These measurements lead to an improved knowledge of the low-$x$ parton
densities which has already been applied in many CR calculations, for example,
calculations of neutrino interaction length, charm production, and
prompt muon yields \cite{Gandhi98a,Pasquali98a}. 

Concerning ultra-high energy CR interactions, it is needed to
extrapolate the measured parton densities
down to lower $x$ by several orders of magnitude. For example, partons
with $x$ values of $10^{-6}$ to $10^{-7}$ dominate minijet and charm
production in proton-air interactions at $E_{\rm lab} \approx 10^{20}$
eV and are important in the case of neutrino-air interactions. The
production of jets with several GeV transverse momentum in very forward
direction involves even smaller $x$ values of about $10^{-10}$.

From the HERA data it is clear that the gluon density rises rapidly at
low $x$. It is convenient to parametrize this in terms of the power
$\Delta_H$ with (see Eq.~(\ref{double-log}))
\begin{equation}
x g(x,Q^2) \sim \frac{1}{x^{\Delta_H}},
\label{singular-gluon}
\end{equation}
where data suggest $\Delta_H \approx 0.25 \dots 0.4$.
However, it is clear from simple geometrical arguments that the rapid
growth of the gluon
density at low $x$ will eventually be tamed by 
saturation effects \cite{Gribov83,Levin90a}. 

\begin{figure}[!htb]
\flushleft{\it democratic distribution:}\\[3mm]
\hspace*{1cm}\epsfig{figure=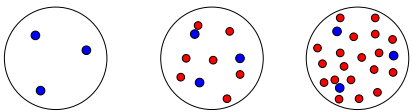,width=5.4cm}
\flushleft{\it hot spots:}\\[3mm]
\hspace*{1cm}\epsfig{figure=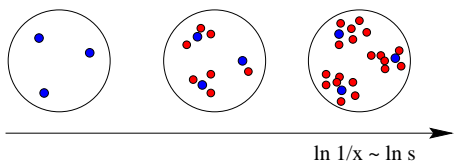,width=6cm}
\vspace*{-5mm}
\caption{Two possible scenarios for the evolution of the parton density
for decreasing $x$.
\label{fig-4a}
}
\end{figure}

Let's consider the
scattering of a quark with the virtuality $Q^2$ off a proton.
This virtual quark probes an effective transverse area of the size of
$\sim \alpha_s^2(Q^2)/Q^2$. As soon as the average transverse distance between 
the gluons in the
proton becomes smaller than $\sim 1/\sqrt{Q^2}$, several gluons
participate in a single quark-proton interaction and the effective
number of gluons "seen" by the quark is smaller than naively expected. 
At low $x$ the effective
number of gluons saturates with an upper limit of 
\begin{equation}
x g(x,Q^2) \sim R_0^2 Q^2,
\label{saturation-limit}
\end{equation}
where $R_0$ is a measure of the radius of the transverse area available
to gluons in the proton. This could be either the proton radius ($R_0^2
\sim 5$ GeV$^{-2}$) itself
or the size of gluon clouds around the valence quarks (hot spot
scenario, $R_0^2 \sim 2$ GeV$^{-2}$), see Fig.~\ref{fig-4a}.
It is important to note that saturation effects might become important
already for $xg(x,Q^2) \ll R_0^2 Q^2$ as diffractive processes
observed in DIS indicate (for example, see \cite{Ahmed94a,Derrick94d}). 
Indeed, there are theoretical arguments supporting the
hypothesis that the gluon density might not reach the saturation limit
(\ref{saturation-limit}) implied by such simple geometrical
considerations  (see, for example,
\cite{Capella94b,Capella96a,Frankfurt96a}; an overview on high-density
parton evolution equations can be found in
\cite{Levin98a}).

\begin{figure}[!htb]
\centerline{\epsfig{figure=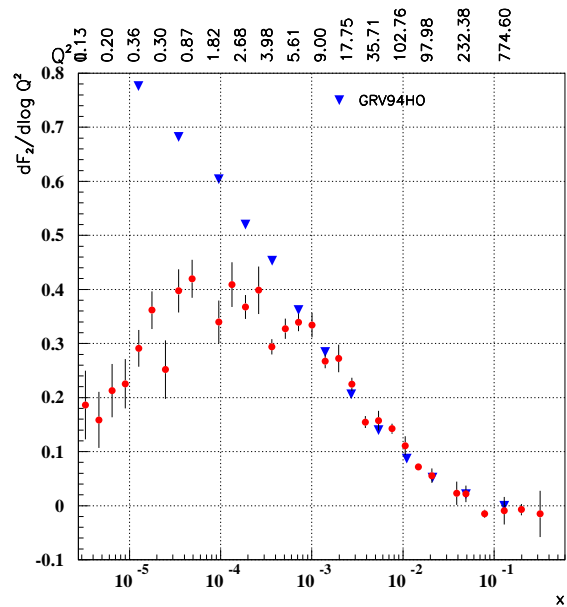,width=7.5cm}}
\vspace*{-5mm}
\caption{$F_2$ scaling violation plotted as function of Bjorken-$x$.
ZEUS data (circles) \protect\cite{Doyle98a} are compared to the 
GRV prediction \protect\cite{Gluck95a} (triangles).
\label{fig-4b}
}
\end{figure}

The gluon density is closely linked to the scaling violation of $F_2$
\begin{equation}
x g(x,Q^2) \simeq \frac{4\pi}{(40/27)\alpha_s}
\frac{\partial F_2(\frac{x}{2},Q^2)}{\partial \ln Q^2}\ .
\label{gluon-scaling-violation}
\end{equation}
This gives a handle to measure saturation
effects. It follows from Eq.~(\ref{saturation-limit}) that in case of
saturation 
\begin{equation}
\frac{\partial F_2}{\partial \ln Q^2} \sim R_0^2 Q^2
\end{equation}
instead of $\partial F_2/\partial \ln Q^2 \sim x^{-\Delta_H}$. In
Fig.~\ref{fig-4b} ZEUS data are shown together with the prediction
according to the DGLAP evolution equations (the Gl\"uck-Reya-Vogt (GRV)
\cite{Gluck95a} parton densities have been used). 
Although it is possible to fit the data within the DGLAP framework
assuming a valence-like gluon density 
at low $x$ \cite{Martin98b,Breitweg98b}, 
it is very likely
that first hints of parton density saturation have been found
\cite{Ellis98a}.

The steep rise and the expected saturation of the gluon density at low
$x$ have several implications for the interpretation of CR
interactions. On one hand it is clear that the minijet cross section
in hadron-air interactions increases rapidly with $\sqrt{s}$. On the
other hand the saturation effects make it very difficult to derive
reliable cross sections for all processes where gluon densities at very
low $x$ are involved. This applies in particular to the minijet and
charm production cross sections in proton-proton and hence in hadron-air
collisions and also, but to a lesser extent, to ultra-high energy neutrino
cross sections.

In most of the hadronic air shower simulation programs the production of
minijets is one important ingredient. These Monte Carlo event generators
are based on the DGLAP evolution equations (\ref{DGLAP}). Therefore it is
interesting to consider the implication of the low-$x$ HERA data for the
high-energy extrapolations done within these models. Without referring to
some particular model, the HERA data imply that
\begin{itemize}
\item
the charged particle multiplicity rises with the energy 
faster than $\ln(s)$ but slower than $s^{\Delta_H}$, and
\item
the mean transverse momentum of hadronic secondaries rises faster than
$\ln(s)$.
\end{itemize}

\begin{figure}[!htb]
\centerline{\epsfig{figure=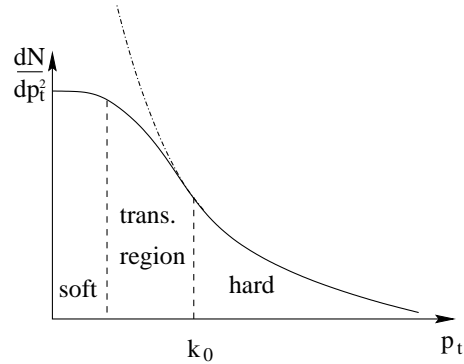,width=6cm}}
\vspace*{-5mm}
\caption{Transverse momentum distribution of hadrons produced in the
central pseudorapidity region in CMS.
\label{fig-5}
}
\end{figure}

\begin{figure}[!htb]
\centerline{\epsfig{figure=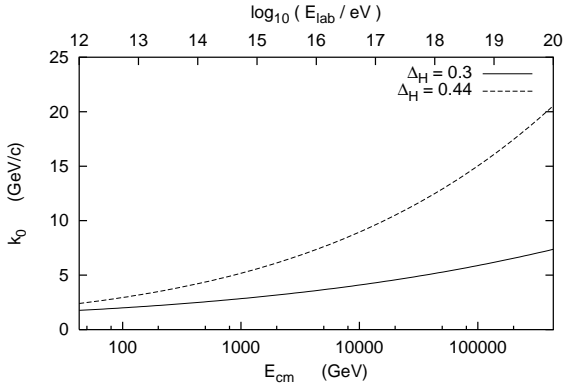,width=7.5cm}}
\vspace*{-5mm}
\caption{Estimate for the parameter $k_0$ characterizing the transverse
momentum below which saturation effects should be important.
\label{fig-6}
}
\end{figure}

This can be understood by considering the transverse momentum
distribution of hadronic secondaries produced in $pp$ interactions
as shown in Fig.~\ref{fig-5}.
The region labeled "hard" is assumed to be calculable on the basis of
the QCD-improved parton model. The "soft" region is dominated by
saturation effects and cannot be calculated perturbatively at all. The
transition region is effected by saturation effects but might still be
calculable with perturbative means. For hard processes the 
differential jet cross section follows from
\begin{equation}
\frac{d\sigma}{d\eta dp_\perp^2} \sim \frac{\alpha^2_s(p_\perp^2)}{p_\perp^4} 
\ x_1 g(x_1,p_\perp^2)\ x_2 g(x_2,p_\perp^2)\ ,
\end{equation}
where $x_1$, $x_2$ are the momentum fractions of the partons engaged in
the hard scattering and $p_\perp$ is the jet transverse momentum and
$\eta$ denotes the pseudorapidity.
After integration over the phase space of the final state particles one
finds, see for example \cite{Kwiecinski91a},
\begin{equation}
\sigma(p_\perp \ge k_0) \sim s^{\Delta_H},
\end{equation}
for a singular gluon distribution (\ref{singular-gluon}).
Consequently, the hard part of the cross section rises much faster with
the energy than the soft part ($\sigma_{\rm soft} \sim s^{\Delta_S}$,
$\Delta_S = 0.07 \dots 0.15$). In order to maintain a steady transition
between soft and hard processes the value of $k_0$ has to increase with
the energy (which is also obvious from Eq.~(\ref{saturation-limit})).
A simple estimate for the parameter $k_0$ 
can be made using \cite{Gribov83}
\begin{equation}
k_0 \sim \exp\left\{c \sqrt{\ln s}\right\}\ ,
\label{k0-estimate}
\end{equation}
which follows from 
Eqs.~(\ref{DGLAP-gluon},\ref{saturation-limit}). The parameter $c$
depends on the assumed low-$x$ behaviour of the gluon density
(\ref{singular-gluon}). In Fig.~\ref{fig-6} the results for $c=0.9$ 
($\Delta_H = 0.3$) and $c = 1.12$ ($\Delta_H = 0.44$) are shown.

In a DGLAP-based model, realizing steeply rising parton densities at low
$x$ but neglecting saturation effects, the average multiplicity of
secondaries becomes proportional to the minijet rate at very high
energies. Consequently the multiplicity increases with $s$ according to a
power law
\begin{equation}
n_{\rm ch} \sim n_{\rm jet} \approx \frac{\sigma_{\rm jet}}{\sigma_{\rm
ine}}
\sim \frac{s^{\Delta_H}}{s^{0.08}} \sim s^{0.1\dots 0.3}\ .
\label{steep-mul}
\end{equation}
The inclusion of saturation effects reduces significantly this rapid
increase
\begin{equation}
n_{\rm ch} \sim k_0^2 \sim \exp\left\{ 2 c \sqrt{\ln s} \right\}
~~~~~\gg~~ \ln(s)
\label{moderate-mul}
\end{equation}
but the multiplicity still grows considerably faster than $\ln(s)$ as
implied by limiting fragmentation and Feynman scaling. 
Hence, HERA data exclude multiplicity
extrapolations of the type (\ref{steep-mul}) as found, for example, in
{\sc QGSjet} \cite{Kalmykov97a} but also disfavour a pure $\ln(s)$ 
extrapolation.
\begin{figure}[!htb]
\centerline{\epsfig{figure=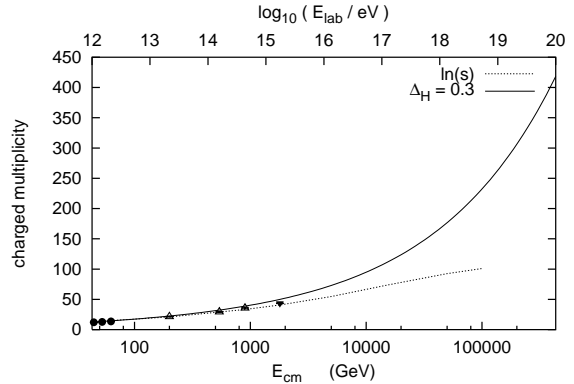,width=7.5cm}}
\vspace*{-5mm}
\caption{Average charged particle multiplicity in $pp$ collisions.
The data are taken from
\protect\cite{Breakstone84b,Ansorge89,Alner85,Lindsey92a}
\label{fig-7a}
}
\end{figure}
In
Fig.~\ref{fig-7a} the two limiting extrapolations are shown, the upper
curve corresponds to a DGLAP implied power law and the lower curve
represents a $\ln(s)$ extrapolation of the collider data.

\begin{figure}[!htb]
\centerline{\epsfig{figure=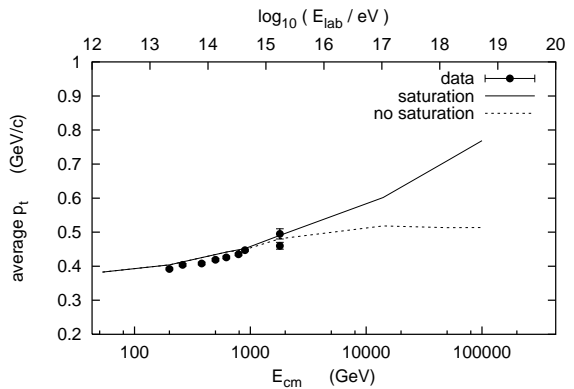,width=7.5cm}}
\vspace*{-5mm}
\caption{Average transverse momentum of secondaries in $pp$ collisions.
Shown are model extrapolations of collider data 
\protect\cite{Albajar90,Abe88}
with and without
taking into account saturation effects. The calculation has been done
with {\sc Phojet} \protect\cite{Engel95a,Engel95d}.
\label{fig-7b}
}
\end{figure}

In the very high energy limit the
mean transverse momentum of secondaries is of the order of $k_0$.
Neglecting saturation effects (which means assuming $k_0=$const.), 
the mean transverse momentum of
secondaries does not increase beyond a certain value even at ultra high
energies.
By contrast, the inclusion of saturation effects leads to an increase of
the average $p_\perp$ which is faster than $\ln(s)$. The size of this
effect can be seen in Fig.~\ref{fig-7b}.
Most of the currently used models show a very moderate increase of the
average $p_\perp$ which is comparable with the dashed curve
\cite{Knapp96a}. Only the {\sc Dpmjet} Monte Carlo \cite{Ranft95a}
predicts a rapid increase
of the average transverse momentum similar to the solid curve in
Fig.~\ref{fig-7b}.
In air shower experiments, the lateral spread of muons with energies 
of the order of 1 GeV might allow to distinguish between the different
model predictions.


\section{Hadronic final state in DIS\label{part4}}

\begin{figure}[!htb]
\centerline{\epsfig{figure=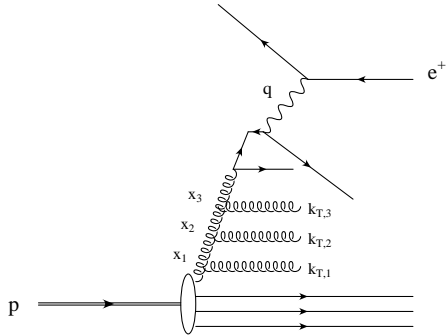,width=6cm}}
\vspace*{-5mm}
\caption{QCD-improved parton model view on DIS.
\label{fig-8a}
}
\end{figure}

\begin{figure}[!htb]
\hspace*{1cm}\begin{turn}{180}
\epsfig{figure=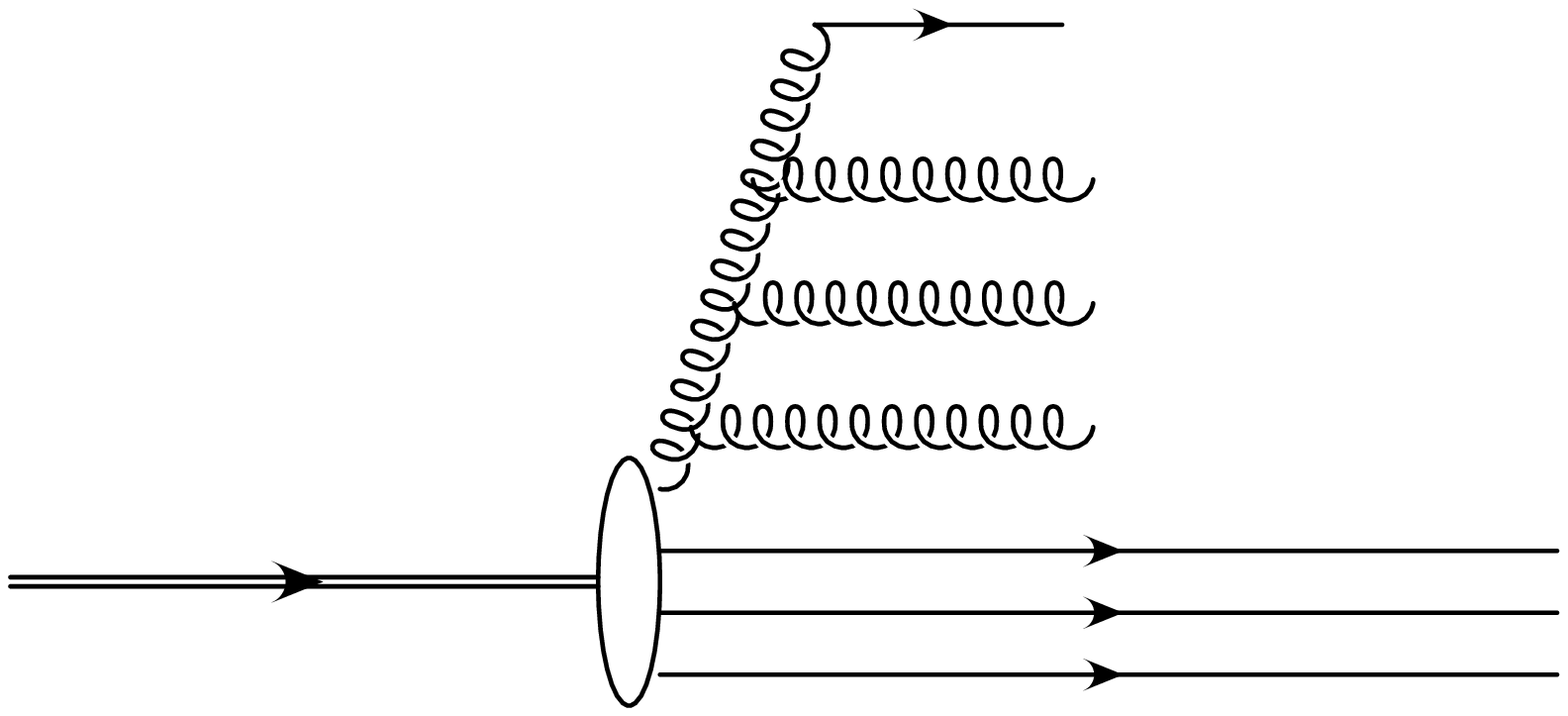,height=2.1cm}
\end{turn}
\vspace*{-5mm}
\epsfig{figure=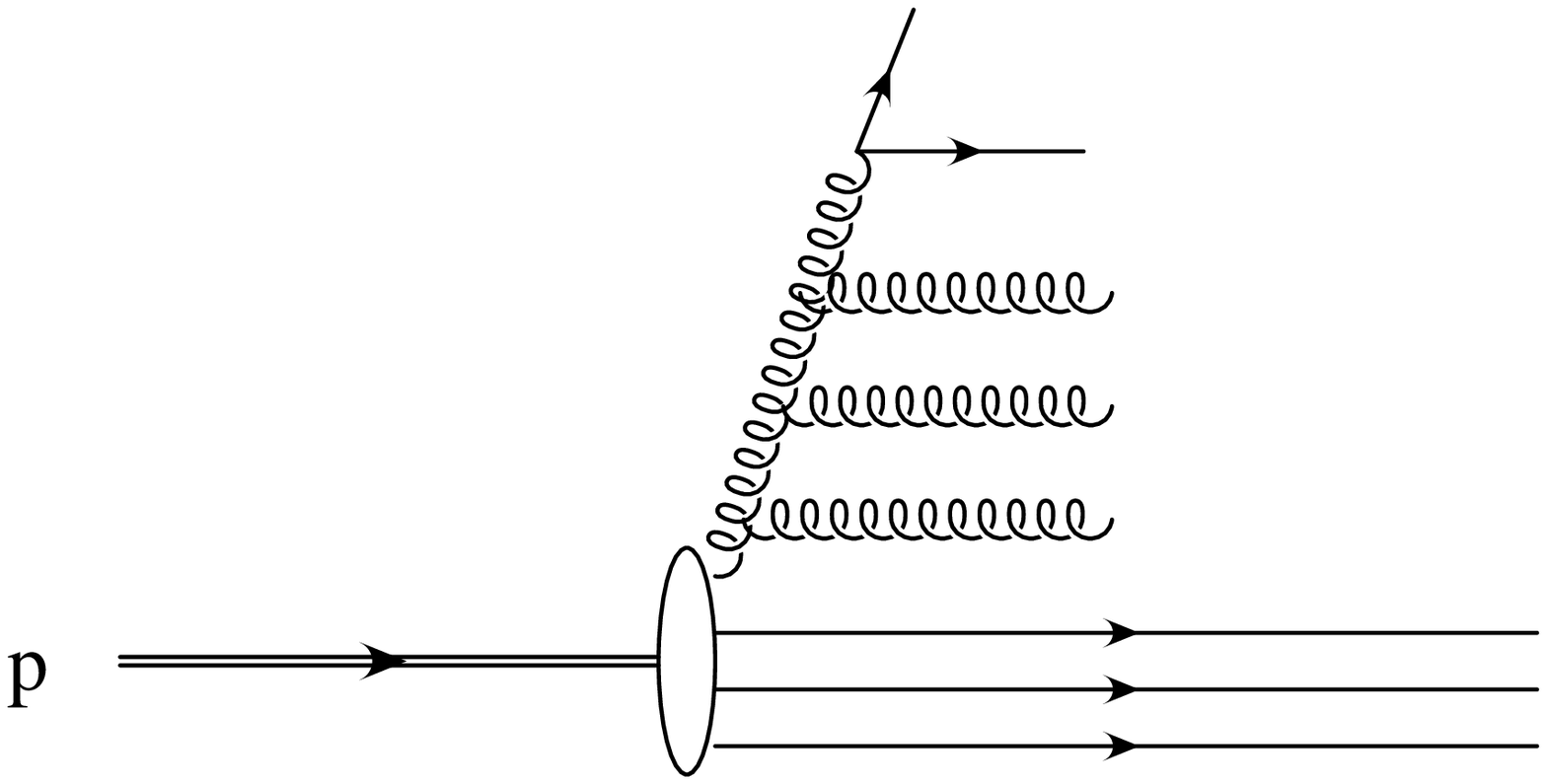,width=5.5cm}
\vspace*{-5mm}
\caption{QCD-improved parton model view on jet production in
proton-proton scattering.
\label{fig-8b}
}
\end{figure}

\begin{figure}[!htb]
\centerline{\epsfig{figure=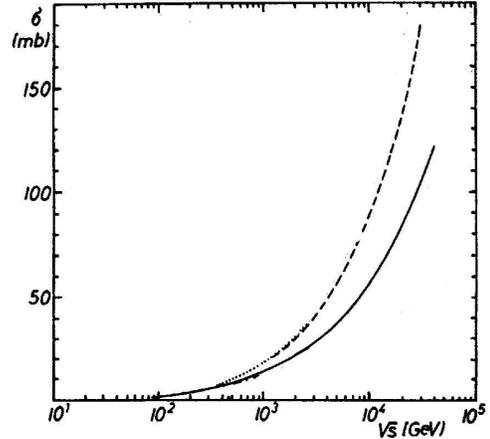,width=6.5cm}}
\vspace*{-5mm}
\caption{Comparison of the hard cross section as calculated within the
collinear factorization approximation (dashed line) and an alternative
approach which is expected to be more suited for low-$x$ processes
(solid line). The graph is taken from \protect\cite{Kwiecinski87a}.
\label{fig-10}
}
\end{figure}

Not only structure function data but
also the knowledge gained at HERA on the hadronic final state of
photon-proton interactions are of interest for CR physics.
For example, hard interaction processes are governed by the same parton
radiation processes in proton-air interactions and in DIS. As can be
seen by comparing Fig.~\ref{fig-8a} with \ref{fig-8b}, the parton
emissions giving rise to the gluon which couples to the quark in the
hard interaction are similar.
The emission of partons (the gluons shown in Figs.~\ref{fig-8a} and
\ref{fig-8b}) is assumed to obey the DGLAP equations. Since basically
almost all perturbative calculations including the event generators used
in CR physics are based on these evolution equations, an experimental
confirmation is very important.\footnote{The description of the $Q^2$-scaling
violation of the structure
function $F_2$ is an experimental confirmation of the DGLAP equations,
however due to the unknown initial conditions not fully conclusive.} 
This is in particular true since there
exist theoretical arguments that there might be a deviation from the
DGLAP predictions for low $x$ processes
\cite{Kuraev76a-e,Kuraev77b-e,Balitsky78b}.
In the case of DIS the kinematics of the quark loop can be fixed by
selecting events with certain scattering angles. 
By means of measuring
the transverse momenta of the gluon-induced jets $k_{T,1}\dots k_{T,3}$
the predictions of (\ref{DGLAP}) can be
tested. The DGLAP evolution equations imply
\begin{eqnarray}
&Q^2& \gg k_{T,3}^2 \gg k_{T,2}^2 \gg k_{T,1}^2 \nonumber\\
& & x_1 \gg x_2 \gg x_3\ .
\label{ordering}
\end{eqnarray}
\begin{figure}[!htb]
\centerline{\epsfig{figure=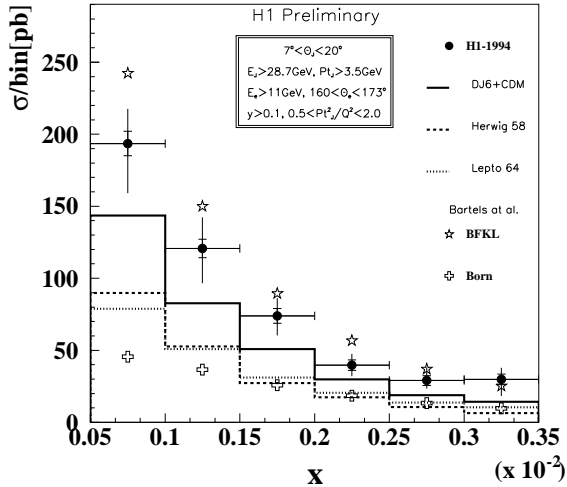,width=7.5cm}}
\vspace*{-5mm}
\caption{Comparison of H1 data on forward-jets with theoretical
predictions. The dashed and dotted curves correspond to DGLAP based
Monte Carlo calculations. The solid curve and the stars represent
alternative approaches. For a detailed discussion see
\protect\cite{Adloff98a}.
\label{fig-9}
}
\end{figure}
In Fig.~\ref{fig-9} the results of the H1 measurement for forward going
jets are shown. Forward-jets roughly correspond to jets produced by
the gluon with $k_{T,1}$ in Fig.~\ref{fig-8a}.
The inclusive cross section for these jets with $0.5 <
k_T^2/Q^2 < 2$ is plotted as function of the Bjorken-$x$ characterizing
the DIS process. 
Clearly the DGLAP based calculations cannot describe the data. Further
measurements are needed to confirm this deviation. There are
some important consequences if the DGLAP based approximations break down 
at low $x$.
\begin{itemize}
\item
The incoming partons of hard scattering processes have sizable
transverse momenta which cannot be neglected.
\item
Strong $k_T$ ordering (\ref{ordering}) does not hold at low $x$.
\item
One parton emission chain can give rise to more than two minijets having
almost the same transverse momentum.
\end{itemize}
The last conclusion can be mathematically re-phrased to
\begin{equation}
\sigma_{\rm hard} \not = \frac{1}{2} \int
\frac{d\sigma_{\rm jet}}{d^2 p_{\perp,\rm jet}}\ d^2 p_{\perp,\rm jet}
\ ,
\end{equation}
where $\sigma_{\rm hard}$ is the cross section for hard processes and 
$d\sigma_{\rm jet}/d^2 p_{\perp,\rm jet}$ is the inclusive jet cross
section.
Up to now all event generators applied in CR simulations assume that the
hard cross section (which is needed for the calculation of the total cross 
section) is half the inclusive single-jet cross section given by
perturbative QCD and collinear factorization (DGLAP approach).
The size of the effect has been estimated in \cite{Kwiecinski87a} and is
shown in Fig.~\ref{fig-10}.

Concerning minimum bias physics, photon-proton collisions are very
similar to pion-proton or proton-proton interactions. This is well
understood in terms of the Vector Dominance Model
(see, for example, \cite{Donnachie78-b}). 
The transition form real ($Q^2 \approx 0$) to
highly virtual photons changes the event properties. 
The hadronic final state can be subdivided
into photon fragmentation, central, and proton fragmentation regions.
As has been confirmed experimentally,
hadron production in the proton fragmentation region is almost
independent of the photon virtuality $Q^2$. This supports the hypothesis
of factorization \cite{Bjorken73a} and allows us to compare data on
leading baryon production in $\gamma p$ collisions to Monte Carlo
predictions for $pp$ interactions at the same energy.
\begin{figure}[!htb]
\centerline{\epsfig{figure=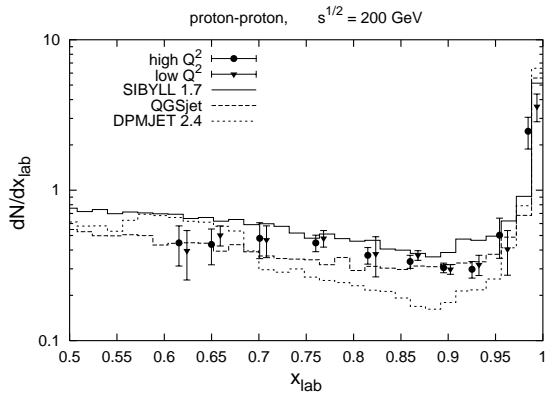,width=7.5cm}}
\vspace*{-5mm}
\caption{Energy fraction $x_{\rm lab}$ carried by the leading proton.
The data are photoproduction and DIS measurements whereas the curves 
are model predictions for $pp$ collisions.
\label{fig-11}
}
\end{figure}
In Fig.~\ref{fig-11} the energy fraction $x_{\rm lab} = 
E_{\rm baryon}/E_{\rm beam}$ carried by the leading proton
in photon-proton and proton-proton collisions is shown
\cite{Garfagnini98a,Schmidke98a}. 
There is no $Q^2$-dependence of the data within 
the experimental errors. The data
show that there is no dip close to $x_{\rm lab} \approx 0.9$ as might be
expected from the naive application of 
Feynman scaling violation arguments. The {\sc Sibyll}
\cite{JEngel92,Fletcher94} and
{\sc QGSjet} \cite{Kalmykov97a} models describe the data reasonably 
well whereas {\sc Dpmjet} \cite{Ranft95a}
seems to underestimate the data close to 
$x_{\rm lab} \approx 0.9$.


\section{Summary and outlook\label{part5}}

HERA data and theoretical work related to HERA physics clearly
improved our understanding of hadron production in such a way that the
uncertainties in the model predictions for very high energy cosmic ray 
interactions can be reduced.

Experimental evidence has been found for a steeply rising gluon density
in the proton at low $x$. The inevitable consequence of this finding is
that saturation effects are important. First experimental signs for this
have been reported. The implication of these results to high-energy
extrapolations done within models for CR interactions include 
\begin{itemize}
\item
the particle multiplicity grows faster than $\ln(s)$ but slower than
$s^{\Delta_H}$
\item
the average transverse momentum of secondaries rises faster than
$\ln(s)$ 
\end{itemize}

Investigations of the hadronic final state in DIS at low $x$ reveal
deviations from the prediction obtained in the "standard" DGLAP
framework (collinear factorization). Up to now, the physics of
hard processes at low $x$ is theoretically not well understood. 
However, it can be expected that the research activity on this subject
triggered by HERA measurements will soon lead to a considerably
improved understanding of  typical low-$x$ processes such as minijet
and charm production.

It has been shown that HERA data
can be used to reduce the uncertainties in Monte Carlo model predictions
concerning central hadron or jet production as well as leading
baryon distributions.

\vspace*{3mm}

\noindent
{\bf Acknowledgments}\\
The author acknowledges the excellent collaboration with T.~Gaisser,
T.~Stanev, and P.~Lipari on the subject of this work.
He is grateful to  J.~Ranft and S.~Roesler for many
discussions.
The work is supported by the
U.S.\ Department of Energy under Grant DE-FG02-91ER40626.





\end{document}